\documentclass[12pt]{article}
\usepackage{scicite}
\usepackage{times}
\usepackage{graphicx}

\newcommand{\ga}{\ifmmode\stackrel{>}{_{\sim}}\else$\stackrel{>}{_{\sim}}$\fi}
\newcommand{\la}{\ifmmode\stackrel{<}{_{\sim}}\else$\stackrel{<}{_{\sim}}$\fi}
\newcommand{\fdg}{\mbox{$.\!\!^\circ$}}

\newcommand{\arcsec}{\hbox{$^{\prime\prime}$}}
\newcommand{\farcm}{\mbox{$.\mkern-4mu^\prime$}}

\newenvironment{sciabstract}{%
\begin{quote} \bf}
{\end{quote}}

\newcounter{lastnote}
\newenvironment{scilastnote}{%
\setcounter{lastnote}{\value{enumiv}}%
\addtocounter{lastnote}{+1}%
\begin{list}%
{\arabic{lastnote}.}
{\setlength{\leftmargin}{.22in}}
{\setlength{\labelsep}{.5em}}}
{\end{list}}

\title{Twenty-One Millisecond Pulsars in Terzan~5 Using the Green Bank
  Telescope}

\author
{Scott M.~Ransom,$^{1,2\ast}$ Jason W.~T.~Hessels,$^{2}$ Ingrid H.~Stairs,$^{3}$\\ Paulo C.~C.~Freire,$^{4}$ Fernando Camilo,$^{5}$ Victoria M.~Kaspi,$^{2}$ \\and David L.~Kaplan$^{6}$\\
\\
\normalsize{$^{1}$NRAO, 520 Edgemont Rd., Charlottesville, VA 22903, USA}\\
\normalsize{$^{2}$Department of Physics, McGill University, Montreal, QC H3A 2T8, Canada}\\
\normalsize{$^{3}$Department of Physics and Astronomy, University of British Columbia,} \\
\normalsize{6224 Agricultural Road, Vancouver, BC V6T 1Z1, Canada}\\
\normalsize{$^{4}$NAIC, Arecibo Observatory, HC03 Box 53995, PR 00612, USA}\\
\normalsize{$^{5}$Columbia Astrophysics Laboratory, Columbia University, 550 West 120th Street,}\\
\normalsize{New York, NY 10027, USA}\\
\normalsize{$^{6}$Center for Space Research, Massachusetts Institute of Technology,}\\
\normalsize{70 Vassar Street, Cambridge, MA 02139, USA}\\
\normalsize{$^\ast$To whom correspondence should be addressed; E-mail: sransom@nrao.edu.}
}

\date{}

\begin{document} 
\baselineskip16pt
\maketitle 

\begin{sciabstract}
  We have discovered 21 millisecond pulsars (MSPs) in the globular
  cluster Terzan~5 using the Green Bank Telescope, bringing the total
  of known MSPs in Terzan~5 to 24.  These discoveries confirm
  fundamental predictions of globular cluster and binary system
  evolution.  Thirteen of the new MSPs are in binaries, of which two
  show eclipses and two have highly eccentric orbits.  The
  relativistic periastron advance for the two eccentric systems
  indicates that at least one of these pulsars has a mass
  $>$1.68\,M$_{\odot}$ at 95\% confidence. Such large neutron star
  masses constrain the equation of state of matter at or beyond the
  nuclear equilibrium density.
\end{sciabstract}

The extremely high stellar densities (10$^4$$-$10$^6$\,pc$^{-3}$) in
the cores of globular clusters (GCs) result in stellar interactions
which produce and destroy binary systems as well as exchange their
members \cite{mh97}.  Formed by the death of massive stars early in a
cluster's history, neutron stars (NSs) usually reside near the cores
of clusters due to their relatively large masses and the mass
segregation induced by dynamical friction.  There they are likely to
interact with one or more stars over the 10$^{10}$\,yr lifetimes of
GCs.  These interactions lead to a production rate of low-mass x-ray
binaries (LMXBs) and their progeny such as MSPs [via the recycling
mechanism \cite{acrs82}], that is highly enhanced compared to the rate
in the Milky Way Galaxy. Before our observations, there were 80
pulsars (most of them binary MSPs) known in 24 GCs \cite{pauloweb},
with the relatively massive and nearby cluster 47~Tucanae containing
22 of these \cite{clf+00}.  Finding and monitoring many pulsars in a
single cluster provides unique probes into a range of GC, binary
evolutionary, and stellar astrophysics \cite{and92,fck+03}.  In
addition, GCs produce exotic systems such as highly eccentric
\cite{fgri04} and MSP$-$main sequence star binaries \cite{dpm+01a}.

The interaction rate between NSs and other stars or binaries in a GC
is a complex function of the total cluster mass, the size of the
cluster core \cite{core} and its stellar density, the initial stellar
mass function, and the level of mass segregation present in the
cluster \cite{ver03}.  However, relatively simple theoretical
modelling of stellar interaction rates \cite{ver02}, as well as the
known LMXB and several additional x-ray sources detected recently with
the {\em Chandra X-ray Observatory} \cite{heg+03} indicate that the
dense, massive, and metal-rich GC Terzan~5 has one of the highest
stellar interaction rates of any cluster in the Galaxy \cite{pla+03}
and perhaps also the largest number of MSPs \cite{sp95,ka96}.  But
because Terzan~5 [Galactic coordinates ($l$,$b$)=(3\fdg8, 1\fdg7)] is
distant ($D$=8.7$\pm$2\,kpc), and located within $\sim$1\,kpc of the
Galactic center \cite{clge02}, the large column density of
interstellar free electrons (i.e.~the dispersion measure,
DM$\sim$240\,pc\,cm$^{-3}$) produces considerable dispersive smearing
($\propto \nu^{-3}$ for typical pulsar search data) and scatter
broadening ($\propto \nu^{-4.4}$) of radio pulses, which hinders
pulsation searches for MSPs at the often observed radio frequencies of
$\nu$=400$-$1400\,MHz.

Deep radio images of Terzan~5, which are not affected by the
dispersive effects of the interstellar medium (ISM), were made with
the Very Large Array (VLA) and showed what appeared to be the
integrated emission from many tens or even hundreds of pulsars with
low flux densities \cite{fg00}.  Surprisingly, though, numerous deep
searches for pulsars over the past 15 years using the Parkes radio
telescope at 1400\,MHz have identified only three \cite{ter5b}: the
11-ms Ter~5~A with its unusual eclipses \cite{lmd+90}, and the two
isolated MSPs Ter~5~C \cite{lmb+00} and D \cite{ran01}.

\paragraph*{Observations and Data Analysis} 
On 17 July 2004, we observed Terzan~5 for 5.9\,hr using the National
Radio Astronomy Observatory's \cite{nrao} 100-m Green Bank Telescope
(GBT) with the Pulsar Spigot backend \cite{kel+05}.  The S-band
receiver provided 600-MHz (1650$-$2250\,MHz) of relatively
interference-free bandwidth in two orthogonal polarizations, which the
Spigot summed and synthesized into 768$\times$0.78125\,MHz frequency
channels every 81.92\,$\mu$s.  With the known DM and scattering time
scale \cite{nt92} toward the cluster, the effective time resolution of
the data was $\sim$0.3\,ms.  We observed only a single position since
the 6$\farcm$5 telescope beam width at these frequencies is
significantly larger than the 0\farcm83 half-mass radius of the
cluster \cite{har96}.  The large unblocked aperture of the GBT, the
wide bandwidth provided by the Spigot and S-band receiver, and the
move to higher observing frequencies together provided an increase in
sensitivity over the Parkes searches \cite{ran01} by factors of
$\sim$5 for typical recycled pulsars and by $\ga$10 for MSPs with spin
periods $P_{\rm psr}$$\la$2\,ms \cite{spec}.

We searched the observation by de-dispersing the raw data into 40
separate time series with DMs ranging from 230$-$250\,pc\,cm$^{-3}$
and spaced by 0.5\,pc\,cm$^{-3}$.  We Fourier-transformed the full
5.9-hr time series as well as 10-, 20- and 60-min sections and
searched them using Fourier-domain acceleration search techniques
\cite{rem02} in a manner similar to that described in Ransom et
al.~\cite{rsb+04} in order to maintain sensitivity to pulsars in
compact binary systems.  In that first observation we discovered 14
new pulsars, Ter~5~E$-$R \cite{ter5e}.  Using eight more observations
taken between July and November 2004, we found seven additional
pulsars and determined the basic orbital parameters of 10 of the 13
new binaries (Fig.~1 and Table~1).  In general, the uncalibrated flux
densities of the pulsars (as computed by comparing the integrated
signal from a pulsar to the predicted total system noise level) were
constant to within a level of $\la$50\% between these observations
implying that diffractive scintillation will not largely affect pulsar
measurements for Terzan~5 at these observing frequencies.  Several
additional binary pulsar candidates from these data remain unconfirmed
but may reappear in future observations at more fortuitous orbital
phases.  Since the volume enclosed by the GBT beam at 1950\,MHz out to
a distance of 10\,kpc is $\sim$10$^{-3}$\,kpc$^3$ and it has been
estimated \cite{lml+98} that there are roughly 10$^5$ observable MSPs
within the Galaxy (with volume $\sim$10$^2$$-$10$^3$\,kpc$^3$), it is
possible that a detectable foreground MSP is within our beam.
However, given that the DM towards Terzan~5 is known to within
$\sim$5\%, the probability of that MSP having a similar DM is much
smaller, and hence we are confident that all of the new pulsars are
members of Terzan~5.  Positions with $\la$1\arcsec\ accuracy of the new
MSPs from pulsar timing will solidify the associations.

\paragraph*{The New Pulsar Population}
With our initial discovery data, the ensemble of pulsars all have
measured DMs with errors $\la$0.1\,pc\,cm$^{-3}$, flux densities
$S_{1950}$ with fractional errors of $\sim$30\%, and precisely
determined spin periods $P_{\rm psr}$.  These measurements allow us to
compare the new systems with known MSPs such as those in 47~Tuc.  Once
precise positions and pulsar spin period derivatives have been
determined from timing observations over the next year (the positions
of the new pulsars are currently known only to within the 6\farcm5
primary beam of the GBT), a variety of tests of the gravitational
potential and internal dynamics of Terzan~5 will be possible, in
addition to estimates of the ages and magnetic field strengths of many
of the pulsars \cite{pk94}.

The DMs for the pulsars in Terzan~5 are distributed with a central
Gaussian-like component with an average and standard deviation of
$237.8\pm1.4$\,pc\,cm$^{-3}$, and five outliers having higher and
lower DMs (Table~1).  The overall spread in DM, 9.5\,\,pc\,cm$^{-3}$
(13 times larger than what is observed for 47~Tuc), is the largest
known for any GC and is likely due to the large average DM for
Terzan~5 and irregularities in the ISM along the differing sight-lines
toward its pulsars. The Gaussian-like component of the DM distribution
likely corresponds to a group of centrally concentrated pulsars near
the cluster core.  The position of Ter~5~C, which has an average DM,
is about 10\arcsec\ or one core radius \cite{core} north of the
cluster center and supports such a notion.  Likewise, pulsars with
outlying DMs, such as Ter~5~D and J, are probably more offset like
Ter~5~A, which has a high DM and is located 36\arcsec\ from the center
\cite{lmb+00}.

The VLA radio imaging of Terzan~5 \cite{fg00} revealed numerous point
sources within 30\arcsec\ of the cluster center, as well as
$\sim$2\,mJy of diffuse emission at 1400\,MHz within $\sim$10\arcsec.
The central emission was attributed to 60$-$200 unresolved pulsars
assuming $D$=7.1\,kpc, a standard luminosity distribution, and a
minimum pulsar luminosity ($L_{1400} \equiv S_{1400}\times D^2$) of
$L_{1400,\mathrm{min}}$=0.3\,mJy\,kpc$^2$. A distance to Terzan~5 of
$D$=8.7\,kpc \cite{clge02}, would increase the number of unresolved
pulsars by $\sim$50\%.  For a typical pulsar radio spectral index of
$-$1.6 \cite{lylg95}, and omitting the flux densities of Ter~5~A and C
which are located outside the region of diffuse emission
\cite{lmb+00}, the integrated measured flux density from the other 22
pulsars at 1400\,MHz is $\sim$1.3\,mJy.  If the five pulsars with
outlying DMs reside outside the cluster core, the total flux density
from the remaining pulsars is only $\sim$1\,mJy.  None of the new
pulsars can individually (or in combination with nearby Ter~5~C)
account for the bright (1.42\,mJy) point source ``N'' located
12\arcsec\ north of the cluster center \cite{fg00}.  However, given
its very wide pulse profile, lack of a flat off-pulse baseline, and
positional coincidence \cite{lmb+00}, Ter~5~C could possibly emit
significant unpulsed emission like the MSP J0218$+$4232 \cite{nbf+95}
and thereby account for all of the flux density of source N.

The differential luminosity distribution of the new pulsars resembles
that of the 47~Tuc\cite{clf+00}, M15 \cite{and92}, and the Galactic
disk \cite{lmt85} populations by following the normal $d\log N =
-d\log L$ relation, with no sign of a downturn at the low luminosity
end ($L_{1400}$$\sim$1$-$2\,mJy\,kpc$^2$).  This implies that we have
not reached the lowest luminosities of the intrinsic pulsar
distribution in Terzan~5, nor have we yet reached the sensitivity
limit of the observing system \cite{smin}.  It is likely that tens of
less-luminous pulsars remain to be discovered in Terzan~5, and
possibly even several bright ones (e.g.~the source N).  Such bright
pulsars may remain undetected if they are members of very compact or
massive binaries due to eclipsing or excessive Doppler accelerations
or if they have extremely fast spin periods ($P_{\rm
  psr}$$\ll$1.5\,ms).  Detailed searches using more advanced search
techniques may yet uncover these systems.

The 24 known pulsars in Terzan~5 and the 22 in 47~Tuc appear to have
different spin period distributions. The 47~Tuc pulsars are a
homogeneous population with periods 2.1$-$7.6\,ms \cite{pauloweb},
while those in Terzan~5 have a flatter distribution that includes six
pulsars slower than 7.6\,ms and the four fastest pulsars known in GCs
($P_{\rm psr}$=1.67, 1.73, 2.05, and 2.07\,ms).  A Kolmogorov-Smirnov test
suggests that these samples are drawn from different parent
distributions at 85\% confidence.  These differences may be related to
the dynamical states of the cluster cores [i.e. just pre- or just
post-core collapse for Terzan~5 \cite{clge02}], different epochs of
MSP creation, or the occurrence of unusual evolutionary mechanisms
that only manifest themselves at very high (i.e. $>$10$^5$\,pc$^{-3}$)
stellar densities.

\paragraph*{Individual Pulsars}

The binaries Ter~5~E and W have orbital periods of $P_{\rm
  orb}$$\sim$60\,days and $\sim$4.9\,days, respectively.  In the dense
stellar environment of a GC, such wide binaries have large
cross-sections for stellar encounters which can disrupt them, eject
them from the cluster core, or after multiple collisions, induce
significant eccentricities \cite{rh95} in their initially circular
orbits \cite{pk94}.  For these pulsars, the measured eccentricities
are $e$$\sim$0.02, significantly larger than predicted for binary MSPs
with helium WD companions in the Galaxy \cite{pk94}.  If they are
located near the cluster center, where the stellar densities are
$\ga$$10^5$\,pc$^{-3}$ \cite{har96}, the time scale for interactions
to produce such eccentricities is 10$^8$$-$10$^9$\,yr \cite{rh95},
consistent with the $\ga$10$^9$\,yr ages of most cluster pulsars.
However, the 60-day orbit of Ter~5~E implies a time scale near the low
end of that range, possibly too short for MSP lifetimes.  This may
indicate that it resides further from the center of Terzan~5:
$>$10$^9$\,yr spent in the core could result in many interactions
which would either destroy the binary or induce much larger
eccentricities.  The eccentricities of Ter~5~I and J are too large to
have been produced by this method, especially given the relatively
compact nature of their orbits, indicating that they were formed by a
different mechanism.

At least two of the new binary pulsars, Ter~5~O and P, have been
observed to eclipse, although given the factor of 10 difference in the
inferred companion masses ($m_2$$\ga$0.036\,M$_{\odot}$ for Ter~5~O
and $\ga$0.38\,M$_{\odot}$ for Ter~5~P) and eclipse durations
($\sim$0.05$P_{\rm orb}$ for Ter~5~O and $\ga$0.5$P_{\rm orb}$ for
Ter~5~P), the systems are unalike.  Ter~5~O, the third fastest MSP
known, is similar to the Black-Widow eclipsing MSP B1957$+$20
\cite{fst88} with a very low-mass companion and an eclipse duration
($\sim$15$-$20\,min) corresponding to a physical size
($\sim$0.03\,R$_{\odot}$) much larger than the companion star's Roche
lobe.  Such systems are common in GCs \cite{kdb03}. Ter~5~P, the
fourth fastest MSP known, is unusual and most similar to the binary
MSP NGC~6397~A, whose peculiar red straggler or sub-subgiant companion
causes irregular and long-duration eclipses of the radio pulses
\cite{dpm+01a}.  Ter~5~P's eclipses are also irregular and correspond
to an eclipsing region of several solar radii in size, indicating the
companion is likely a peculiar evolved star as well.  If so, the
system may have been created after at least one exchange encounter
where the MSP's original companion (which had spun up the pulsar) was
ejected and replaced by a main sequence star.  Such an encounter might
eject the system from the core of the cluster as has been observed for
NGC~6397~A [although alternative formation scenarios for this pulsar
have been proposed as well \cite{pdc+05}].

The pulsars Ter~5~I and J are in highly eccentric orbits with
companions of at least 0.24 and 0.38\,M$_{\odot}$, respectively
(Fig.~2).  The companions are not sufficiently massive to be NSs
unless the systems have improbably low inclination angles $i$ (i.e.
are almost face-on).  If they were main-sequence or giant stars,
eclipses would likely be observed for a variety of inclinations given
the compactness of the orbits (orbital separations of
$\sim$5R$_{\odot}$).  Additionally, the orbital circularization time
scales would be of order 10$^5$\,yr for both pulsars and so no
eccentricity should currently be observed \cite{tas95}.  The
companions are therefore probably WDs.  However, the pulsar recycling
scenario generates MSP$-$WD binaries in nearly circular orbits due to
tidal interactions during mass-transfer and therefore did not produce
these eccentric systems.  A possible scenario involves the off-center
collision of a NS with a red giant which would disrupt the giant's
envelope and leave the core (now the WD) in an eccentric orbit about
the NS \cite{rs91}.  An alternative scenario involves the exchange of
an initially isolated WD into a binary consisting of an MSP and the
low-mass WD that recycled it.  However, the observed pulsars are only
mildly recycled.  Furthermore, the multiple exchanges required in this
scenario imply that the system was created near the core, where due to
mass segregation the isolated WDs would be more massive than the
observed companions to Ter~5~I and J.  We therefore regard the
collision scenario as more likely.

The high eccentricities of both orbits permitted us to measure their
advances of the angle of periastron,
$\dot\omega=0.255$$\pm$0.001$^\circ$\,yr$^{-1}$ for I and
0.327$\pm$0.004$^\circ$\,yr$^{-1}$ for J, which are likely due to
general-relativistic modification of the Keplerian elliptical orbit.
Under this assumption, the total mass can be derived for each system:
2.17$\pm$0.02\,M$_{\odot}$ for I and 2.20$\pm$0.04\,M$_{\odot}$ for J.
Together with the measured Keplerian mass functions, in both cases
this implies that the most likely NS masses are $>$1.7\,M$_{\odot}$
(Fig.~2).  However, classical contributions to $\dot \omega$ from
tidal or rotational quadrupole mass moments in the companion star must
also be considered \cite{sb76,wex98}.  Tidal deformation is
insignificant for a WD companion \cite{sb76}, but rotationally induced
quadrupoles are possible if the WD is rapidly rotating, as there is no
reason to expect the spin axis of the companion to be aligned with the
orbital angular momentum.

The predicted contributions to $\dot \omega$ for Ter~5~I and J due to
a rotationally induced quadrupole are $\dot \omega_{\rm
  rot}$$\sim$0.01$-$0.02$^{\circ}$\,yr$^{-1}$ times angular and
stellar structure factors \cite{sb76,wex98,sna+02}.  One of the
stellar structure factors, $\alpha_6$, can range up to 15 for WDs
\cite{sb76}, making $\dot \omega_{\rm rot}$ a potentially significant
contribution to the measured $\dot \omega$.  A significant $\dot
\omega_{\rm rot}$, however, will produce changes in $i$ and hence in
the projected semi-major axis $x\equiv a_1 \sin (i)/c$.  The size of
this effect may be written as $\dot \omega_{\rm rot} \sim \dot x_{\rm
  rot} / x$ times a trigonometric factor usually of order unity that
depends on $i$, the angle between the WD rotation axis and the orbital
angular momentum vector, and the phase of the orbital precession
\cite{wex98}. For random choices of these angles the trigonometric
factor is $<$10, $\sim$80\% of the time. By incorporating detections
from Parkes search observations taken in 1998 and 2000 \cite{ran01}
using the now known orbital ephemerides, we have upper limits on $\dot
x$ which imply that $\dot \omega_{\rm rot}$ is
$\la$0.003$^{\circ}$\,yr$^{-1}$ times the trigonometric factor for
both systems. The magnitudes of these rotational-quadrupole
contributions are comparable to our current measurement uncertainties
for $\dot \omega$.  We emphasize that this result does not depend on
the type of companion and is equally valid for main-sequence stars and
WDs.  Therefore, unless the orientations of the orbital and rotational
angular momentum vectors are fine-tuned for both systems, the $\dot
\omega$ measurements are well-described by general relativity.  Under
this assumption, one or both of these NSs are likely to be
unprecedentedly massive: calculation of the joint probabilities
(Fig.~2) indicates that at least one of the pulsars is more massive
than 1.48, 1.68, or 1.74\,M$_{\odot}$ at 99\%, 95\% and 90\%
confidence levels, respectively. A strict upper limit to the masses of
both pulsars is 1.96\,M$_{\odot}$. Similar but slightly less stringent
evidence from pulsar timing for a massive NS in the Galactic low-mass
WD$-$MSP binary J0751$+$1807 has been presented \cite{nss05}. It is an
intriguing question whether the large masses for Ter~5~I and J
resulted from the collision formation mechanism, possibly during a
short period of hypercritical accretion which partially or completely
spun-up the pulsars \cite{rs91}.  GBT observations over the next
1$-$3\,yr will measure the varying delays due to gravitational
redshift and time dilation as Ter~5~I moves in its orbit, which
together are known as $\gamma$ (the measurement of $\gamma$ for
Ter~5~J will take 5$-$10\,yr due to the relatively slow spin period of
the pulsar).  This parameter, along with general relativity and the
already well measured $\dot \omega$, will provide a precise mass for
the pulsar and will likely rule out several soft equations of state
for matter at nuclear densities \cite{lp04}.


\begin{thebibliography}{10}
\bibitem{core}
The core radius is defined as that where the optical surface brightness is half
  its central value.
\bibitem{ter5b}
The 442\,ms pulsar J1748$-$2444 was initially identified as Ter~5~B
  (B1744$-$24B), but is now known to be a foreground pulsar unrelated to
  Terzan~5 \cite{lmb+00}.
\bibitem{ter5e}
Ter~5~E was a candidate (i.e. unconfirmed) pulsar in \cite{ran01}.
\end{thebibliography}

\begin{scilastnote}
\item We thank Fred Rasio, Stein Sigurdsson, and Marten van Kerkwijk
  for extremely useful discussions. Thanks also go to Jim Herrnstein,
  Lincoln Greenhill, Dick Manchester, Andrew Lyne, and Nichi D'Amico
  for providing or aiding with the Parkes data from 1998 and 2000.
  JWTH is an NSERC PGS-D fellow. IHS holds an NSERC UFA and is
  supported by a Discovery grant and UBC start-up funds. FC thanks
  support from the NSF. VMK holds a Canada Research Chair and is
  supported by an NSERC Discovery Grant and Steacie Fellowship
  Supplement, by the FQRNT and CIAR, and by a New Opportunities Grant
  from the Canada Foundation for Innovation.  DLK is a Pappalardo
  Fellow.
\end{scilastnote}

\clearpage

\begin{table}
\begin{center}
\begin{tabular}{cr@{.}lccr@{.}lccc}
\hline \hline
     & \multicolumn{2}{c}{}          & Dispersion  & 1950\,MHz    & \multicolumn{2}{c}{} &         &      &  Minimum \\
     & \multicolumn{2}{c}{$P_{\rm psr}$} & Measure     & Flux Density & \multicolumn{2}{c}{$P_{\rm orb}$} & $x$   &      & $m_2$  \\
PSR  & \multicolumn{2}{c}{(ms)}      & (pc\,cm$^{-3}$) & ($\mu$Jy)    & \multicolumn{2}{c}{(days)}      & (lt-s)  & Eccentricity & (M$_{\odot}$) \\
\hline \hline
A$^e$ & 11&56315 & 242.44 & 1020 & 0&0756  & 0.120 & 0 & 0.089 \\
C & 8&43610  & 237.14 & 360  & \multicolumn{2}{c}{} & & & \\
D & 4&71398  & 243.83 & 41   & \multicolumn{2}{c}{} & & & \\
E & 2&19780  & 236.84 & 48   & 60&06  & 23.6 & $\sim$0.02 & 0.22 \\
F & 5&54014  & 239.18 & 35   & \multicolumn{2}{c}{} & & & \\
G & 21&67187 & 237.57 & 15   & \multicolumn{2}{c}{} & & & \\
H & 4&92589  & 238.13 & 15   & \multicolumn{2}{c}{} & & & \\
I & 9&57019  & 238.73 & 29   & 1&328 & 1.818 & 0.428 & 0.24 \\
J & 80&33793 & 234.35 & 19   & 1&102 & 2.454 & 0.350 & 0.38 \\
K & 2&96965  & 234.81 & 40   & \multicolumn{2}{c}{} & & & \\
L & 2&24470  & 237.74 & 41   & \multicolumn{2}{c}{} & & & \\
M & 3&56957  & 238.65 & 33   & 0&4431 & 0.596 & 0 & 0.14 \\
N & 8&66690  & 238.47 & 55   & 0&3855 & 1.619 & 0.000045 & 0.48 \\
O$^e$ & 1&67663  & 236.38 & 120  & 0&2595 & 0.112 & 0 & 0.036 \\
P$^e$ & 1&72862  & 238.79 & 77   & 0&3626 & 1.272 & 0 & 0.38 \\
Q & 2&812    & 234.50 & 27   & \multicolumn{2}{c}{$>$1?} & Unk. & Unk. & Unk. \\
R & 5&02854  & 237.60 & 12   & \multicolumn{2}{c}{} & & & \\
S & 6&11664  & 236.26 & 18   & \multicolumn{2}{c}{} & & & \\
T & 7&08491  & 237.70 & 20   & \multicolumn{2}{c}{} & & & \\
U & 3&289    & 235.50 & 16   & \multicolumn{2}{c}{$>$1?} & Unk. & Unk. & Unk. \\
V & 2&07251  & 239.11 & 71   & 0&5036 & 0.567 & 0 & 0.12 \\
W & 4&20518  & 239.14 & 22   & 4&877 & 5.869 & 0.015 & 0.30 \\
X & 2&999    & 240.03 & 18   & \multicolumn{2}{c}{$>$1?} & Unk. & Unk. & Unk. \\
Y & 2&04816  & 239.11 & 16   & 1&17 & 1.16 & 0 & 0.14 \\
\hline \hline
\end{tabular}
\caption{
  Known pulsars in Terzan~5 \cite{ter5b}.  Pulsars listed without
  orbital parameters are likely isolated systems while those marked
  with an $^e$ are eclipsing systems.  The errors on the dispersion
  measures (DMs) range from 0.01$-$0.1\,pc\,cm$^{-3}$ and the errors
  on the measured flux densities are $\sim$30\%. The flux densities
  for the eclipsing pulsars include only the times when the pulsar is
  not eclipsed.  The light travel time across the projected pulsar
  semi-major axis is defined as $x\equiv a_1 \sin (i)/c$.
  Eccentricities listed as ``0'' are too small to measure at present
  and have been set to zero for orbital parameter fitting.  The
  minimum companion mass $m_2$ was calculated assuming a pulsar mass
  $m_1$ of 1.4\,M$_{\odot}$ and $i$$=$90$^\circ$ except for Ter~5~I
  and J (Fig.~2).  All measured parameters were determined using the
  {\tt TEMPO} software package \cite{tempo}.}
\end{center}
\end{table}

\clearpage

\begin{figure}[t]
  \begin{center}
    \includegraphics[height=6in,angle=270]{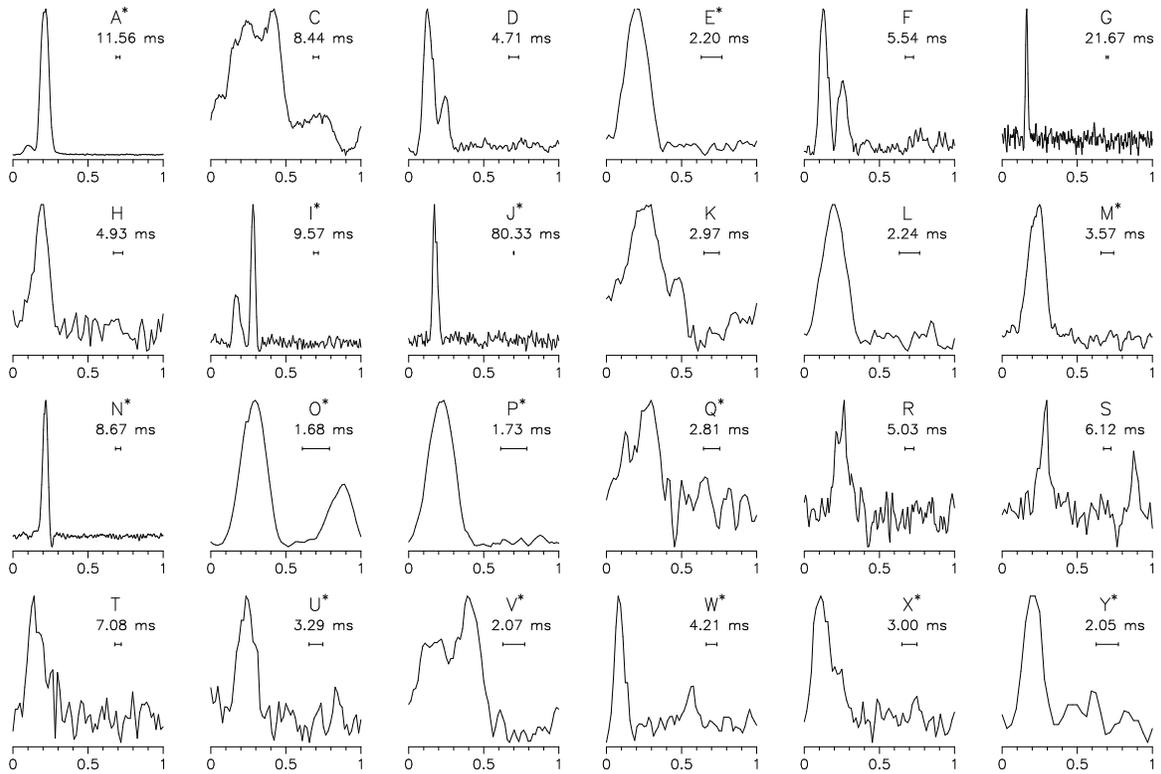}
  \end{center}
  \caption{The 1950\,MHz GBT$+$Spigot pulse profiles for each of the pulsars
    known in Terzan~5.  All but Ter~5~A, C, and D are newly discovered
    \cite{ter5b}.  Each profile is the weighted average of the best
    detections of that pulsar and is a measure of the relative flux
    density as a function of rotational phase.  Asterisks indicate
    that the pulsar is the member of a binary system, and the length
    of the horizontal error bar (0.3\,ms) is the effective system time
    resolution.}
\end{figure}

\clearpage

\begin{figure}[t]
  \begin{center}
    \includegraphics[height=3.1in,angle=0]{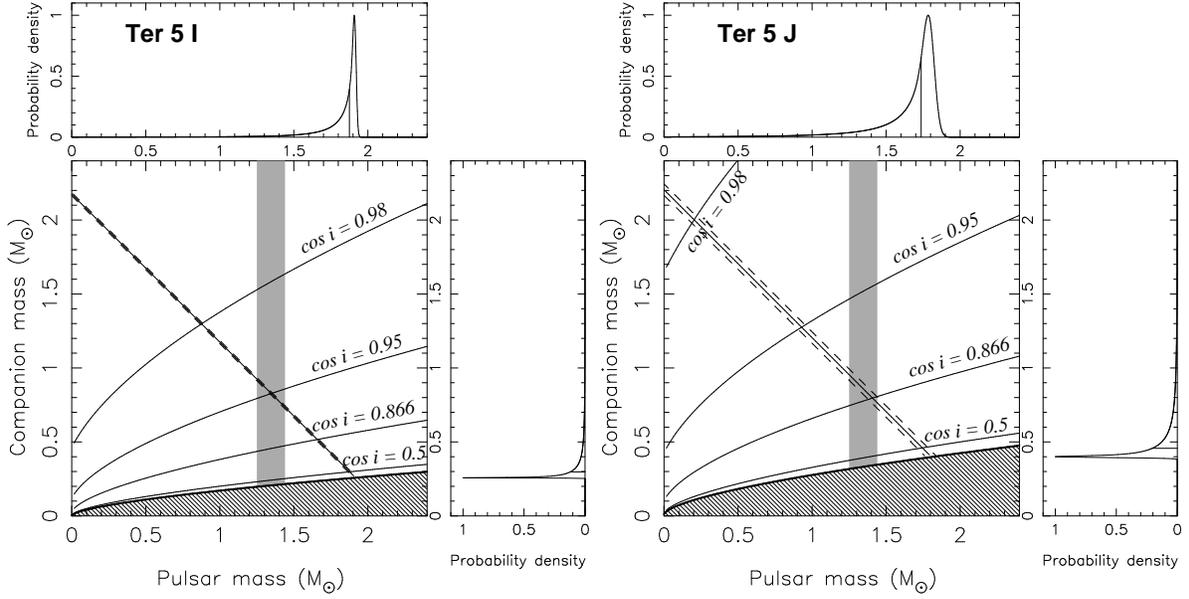}
  \end{center}
  \caption{Pulsar mass vs. companion mass diagrams for the two highly eccentric
    binary pulsars Ter~5~I (left) and Ter~5~J (right).  The hatched
    regions are excluded due to the definition of the Keplerian mass
    function (i.e. $\sin i \le 1$). The diagonal band in the center of
    the figure shows the total system mass with 1-$\sigma$ confidence
    intervals as measured by the general-relativistic advance of
    periastron $\dot\omega$=0.255$\pm$0.001$^\circ$\,yr$^{-1}$ for I
    and 0.327$\pm$0.004$^\circ$\,yr$^{-1}$ for J.  The marginal
    distributions show the masses of the pulsars and companions
    assuming a random distribution of inclinations (i.e., with
    probability density flat in $\cos i$).  The solid curves in the
    main plot indicate inclinations of (from top to bottom)
    11.4$^\circ$, 18.2$^\circ$, 30$^\circ$, 60$^\circ$, and
    90$^\circ$.  The grey vertical band shows the range of precisely
    measured NS masses from relativistic binary radio pulsars
    \cite{lp04}.  In both cases the median pulsar mass (indicated by
    the vertical line in the marginal distribution) lies significantly
    above 1.7\,M$_\odot$, implying that one or both of these pulsars
    is considerably more massive than the NSs that have been well
    measured to date. A strict upper limit to the masses of both
    pulsars is 1.96\,M$_{\odot}$.}
\end{figure}

\end{document}